\documentclass[%
 reprint,
 amsmath,amssymb,
 aps,
]{revtex4-1}

\usepackage[utf8]{inputenc}
\usepackage{tensor, xcolor}
\usepackage{comment, graphicx}

\usepackage[colorinlistoftodos]{todonotes}

\begin{document}

\title{Comment on `Do electromagnetic waves always propagate along null geodesics?'}


\author{Niels Linnemann}
\email{niels.linnemann@uni-bremen.de}
\affiliation{
University of Bremen\\
28359 Bremen,  Germany}\altaffiliation[Also at ]{Rotman Institute of Philosophy, University of Western Ontario, Canada.}
\author{James Read}
\email{james.read@pmb.ox.ac.uk}
\affiliation{%
 Pembroke College, University of Oxford \\
 Oxford OX1 1DW, United Kingdom}%
\vspace{-0.5cm}





\date{\today}

\begin{abstract}
We study the propagation of Maxwellian electromagnetic waves in curved spacetimes in terms of the appropriate geometrical optics limit, notions of signal speed, and minimal coupling prescription from Maxwellian theory in flat spacetime. In the course of this, we counter a recent major claim by Asenjo and Hojman (2017) to the effect that the geometrical optics limit is partly ill-defined in Gödel spacetime; we thereby dissolve the present tension concerning established results on wave propagation and the optical limit.
\end{abstract}

\maketitle

\section{Introduction}

The conventional wisdom that light in relativistic spacetimes propagates on null geodesics can be substantiated in several ways. First: via the observation that light, in the geometrical optics limit, propagates on null geodesics \cite[p.~571]{MTW}.
Second: via the recent technical result that small bodies constructed from Maxwell fields in curved spacetimes `track' null geodesics \cite{GW}. Third: via the core finding that solutions to the Maxwellian wave equation in curved spacetimes have an idealised signal speed of exactly $c$ \cite{Friedlander}. A less well-known aspect of our understanding of the propagation of electromagnetic waves in relativistic spacetimes is that this propagation is associated with a `tail', which moves behind the wavefront at a speed less than $c$ \cite{DB, Faraoni}. Insofar as this tail becomes more pronounced with an increase in gravitational field strength \cite{EllisSciama}, tail production can effectively retard the wavefront to a subluminal speed from a metrological point of view: a detector only registers an incoming wave above a certain threshold; if the power transmission linked to the tail becomes significantly more relevant than that of the actual wavefront, the detector will fire later than otherwise expected.




The recent works \cite{AH, AH2} seek to further develop our understanding of the propagation of light in curved spacetimes. In \cite{AH2}, it is shown that, before taking the geometrical optics limit, light rays in generic spacetimes do not propagate on null geodesics, in the sense that the dispersion relation $K^\mu K_\mu \neq 0$, where $K^\mu$ is the wave vector. In \cite{AH}, it is argued in addition that at least in the G\"{o}del solution, one cannot consistently take the geometrical optics limit (in which, as already mentioned above, the null propagation of light obtains).

These papers certainly constitute advances in our understanding of the propagation of light in curved spacetimes. However, there remains much to be done in analysing their results. This article seeks to clarify comprehensively both the sense in which the claims made in \cite{AH, AH2} are correct, and can be bolstered, as well as the sense in which the claims made in those articles are incorrect, and should be rejected.
More specifically, in this paper we achieve the following tasks: (A) Numerically solving the differential equations describing the propagation of light in curved spacetimes presented in \cite{AH, AH2}, in order to secure a better quantitative grasp of how the dispersion relation (and thereby also the phase and group velocity) of such waves can differ from that associated with null propagation (in this sense, we concur with \cite{AH, AH2}, and take our results to corroborate their claims). (B) Understanding how different curvature couplings in matter dynamics can affect the propagation of light, and thereby lead to distinct physics (again, in this sense our results are consistent with, and develop further, \cite{AH, AH2}). (C) Clarifying exactly when one can and cannot take the geometrical optics limit: in this regard, we find the results presented in \cite{AH} to rest on a questionable limit procedure; we instead side with the orthodoxy in concluding that the geometrical optics limit can invariably be taken consistently, and moreover that both group velocity and the causally relevant signalling speed are $c$ in this limit. (D) Assessing the implications of the above work for the possibility of super- or subluminal wavefront velocities, bearing in mind classic work such as \cite{Brillouin}, and associated more recent discussions (this discussion is important, because explicit consideration of such different notions of wave velocities is absent from \cite{AH, AH2}).

Overall, then, our purpose is to provide an exhaustive study of the claims made in \cite{AH, AH2} regarding the propagation of light in curved spacetimes. The structure of the paper is as follows. In \S\ref{s3}, we present our numerical results which bolster the aforesaid claims regarding the propagation of light outside of the geometrical optics regime. In \S\ref{s4}, we discuss the correct way to take the geometrical optics limit in curved spacetimes. In \S\ref{s5}, we consider how different notions of wave velocities temper the claims made in \cite{AH, AH2} regarding the possibility of superluminal propagation. Before turning to this work in earnest, however, we must first recall some basic details regarding Maxwell's equations in curved spacetimes.

\section{Curvature-coupled Maxwell equations}\label{s2}

The inhomogeneous Maxwell equation for the vector potential in Minkowski spacetime reads
\begin{equation}\label{A-mink}
    -A\indices{^{\alpha,\mu}_{\mu}}+A\indices{^{\mu}^{,\alpha}_{\mu}}=4\pi J^\alpha.
\end{equation}
On minimal coupling (i.e.,~applying the `comma-to-semicolon' rule), one obtains
\begin{equation}\label{A2}
    -A\indices{^{\alpha;\mu}_{\mu}}+A\indices{^{\mu}^{;\alpha}_{\mu}}=4\pi J^\alpha. 
\end{equation}
As has been well-appreciated since (at least) the elegant discussion of Misner \emph{et al.}~\cite[\S16.2]{MTW}, there arises in the implementation of the minimal coupling procedure the possibility of factor-ordering ambiguities. Note, in particular, that if \eqref{A-mink} is written with its partial derivatives exchanged, then one obtains the following curved-spacetime equation, on minimal coupling:
\begin{equation}\label{A3}
    -A\indices{^{\alpha;\mu}_{\mu}}+A\indices{^{\mu}^{;\alpha}_{\mu}} + R\indices{^{\alpha}_{\mu}}A^\mu=4\pi J^\alpha. 
\end{equation}
Typically, \eqref{A2} is treated as the correct wave equation for $A^\mu$ in curved spacetime; \eqref{A3} is rejected. See, for example, \cite[p.~390]{MTW}, where it is stated that ``Coupling to curvature surely cannot occur without some physical reason.'' In this way, minimal coupling is regarded as a (fallible) heuristic for generating equations of motion for matter fields in a curved spacetime which at a point have the same form as their special relativistic counterparts (and in particular do not feature explicit curvature terms): that is, a fallible heuristic for implementing a form of the `strong equivalence principle' (SEP). As minimal coupling fails to implement unambiguously this SEP with respect to \eqref{A-mink} (it yields both \eqref{A2} and \eqref{A3}), the desired form of the general relativistic equations---namely, \eqref{A2}---has to be selected by fiat. Another (again only) heuristic reason to reject \eqref{A3} is its lack of gauge invariance (there is no \emph{a priori} reason why Maxwell equations in curved spacetime have to be gauge invariant). (Our thanks to Jacob Barandes for discussion on this point.)



(An aside: Consider `standard' minimal coupling, in which one obtains \eqref{A2}. Even here, one must take care with claims that these equations are `locally special relativistic', as per the SEP. After all, at least in the gauge $A\indices{^{\mu}_{; \mu}} = 0$,  \eqref{A2} is equivalent, via 
\begin{equation}
R\indices{^{\alpha}_{\mu}} A ^{\mu} = A\indices{^{\mu}_{;\mu}^{\alpha}} - A\indices{^{\mu}^{;\alpha}_{\mu}},
\end{equation}
to
\begin{equation}
-A\indices{^{\alpha;\mu}_{\mu}} - R\indices{^{\alpha}_{\mu}} A ^{\mu} =4\pi J^\alpha.
\end{equation}
The curvature term in this latter presentation of \eqref{A2} will, again, not vanish at a point.)



In any case, \eqref{A3} is a distinct equation from \eqref{A2}, and there is no reason to expect \emph{ab initio} that the equations will make the same empirical predictions; moreover, there is no logical bar to the correct general relativistic versions of Maxwell's equations being \eqref{A3}. Ultimately, only experiment can adjudicate between these different possibilities. Thus, it is incumbent on us to appreciate the empirical consequences of each. In this article, we show that the differences between \eqref{A2} and \eqref{A3} manifest themselves in the velocity of the propagation of light, at least outside of the geometrical optics regime; in this sense, we develop further the claims made in \cite{AH} in this regard (see \S\ref{s3}).

There is one other point which is worth making here. In addition to the factor-ordering ambiguities associated with minimal coupling, other authors have motivated in other ways consideration of the relations between curvature couplings and wave propagation. For example, Drummond and Hathrell consider one-loop vacuum polarisation contribution to the QED effective action and the additional terms in the resulting equations of motion derived therefrom \cite{DH}. The results which we present in this article are consistent with the motivations of these investigations, insofar as one of our central interests is also the effects that such additional couplings can have upon wave propagation. For recent discussion of such work, see \cite{Shore, Butterfield}. (Note that the putative superluminal signalling in the context of the Drummon-Hathrell action has been called into question for making use of questionable approximations \citep{Shore}. Interestingly, we will express a very similar concern towards the claimed failure of the geometrical optics limit in Gödel spacetime in \cite{AH}.)

\section{Solving the Curvature-Coupled Equations}\label{s3}

With this background in hand, we turn now to a quantitative discussion of the propagation of Maxwellian waves in curved spacetimes; our intention is to build upon the discussion of \cite{AH, AH2} that this propagation need not invariably be null.

The source-free version of \eqref{A2} can be written in terms of partial (rather than covariant) derivatives as
\begin{equation}\label{A2'}
    \partial_\alpha \left[ \sqrt{-g} g^{\alpha\mu} g^{\beta\nu} \left( \partial_\mu A_\nu - \partial_\nu A_\mu\right)\right]=0.
\end{equation}
By making a plane wave ansatz in the $\nabla_\mu A^\mu =0$ gauge, i.e.~$A_{\mu} = \xi_{\mu} (x^\lambda) e^{i S(x^\lambda)}$ where $\xi_{\mu}$ and $S$ are not necessarily constant and represent, respectively, the amplitude and phase of the wave, it is shown in \cite{AH2} that the wave `vector' $K_\mu := \partial_{\mu} S$ obeys $K^\mu K_\mu = \chi$,
where ($\tilde{\xi} := \sqrt{\xi_\mu \xi^\mu}$)
\begin{equation}\label{eq6}
    \chi = \frac{\xi_\beta}{\sqrt{-g}\tilde{\xi}^2} \partial_\alpha \left[ \sqrt{-g} g^{\alpha\mu} g^{\beta\nu}\left(\partial_\mu \xi_\nu - \partial_\nu \xi_\mu \right) \right] .
\end{equation}
(A general solution to the homogeneous, standard minimally coupled Maxwell wave equation in curved spacetime was given in \cite{DB} in the context of the bitensor formalism.) From this, one might already infer that in generic spacetimes and before taking the geometrical optics limit, light does not propagate on null geodesics: as pointed out in \cite{AH2}, \eqref{eq6} cannot in general be solved consistently using a constant amplitude ansatz; this in turn, however, implies that the dispersion relation for the wave is non-trivial. (That a non-constant amplitude leads to a non-trivial dispersion relation holds even in flat spacetime.)

One can generalise these results by writing the source-free version of \eqref{A3} as
\begin{equation}\label{A3'}
    \partial_\alpha \left[ \sqrt{-g} g^{\alpha\mu} g^{\beta\nu} \left( \partial_\mu A_\nu - \partial_\nu A_\mu\right)\right] + z R\indices{^{\beta}_{\mu}}A^{\mu} = 0;
\end{equation}
here, we have included a scalar parameter $z$, allowing us to `tune' the curvature-coupled term, the possibility of which arises due, for example, to the factor-ordering ambiguities in the minimal coupling scheme, as discussed in \S\ref{s2}. 

As we will now argue, including such a curvature-coupled term leads to a further change in the dispersion relation. For this, we seek to solve \eqref{A3'} in particular spacetimes. Selecting the FLRW spacetime (as also considered by \cite{AH2} in the case of $z=0$) and using spherical coordinates $(t, r, \phi, \theta)$, the non-trivial components of the metric are
\begin{align}
g_{tt} &= -1, \\
g_{rr} &= \frac{R^2(t)}{1- k r^2}, \\
g_{\phi \phi} &= R^2(t) r^2  \sin^2(\theta), \\
g_{\theta \theta} &= R^2(t) r^2.
\end{align}
Then, making the ansatz $A_\mu = A_{\phi} (r, t) \delta_{\mu \phi}$ (which is slightly more restrictive than that used by \cite{AH2} in their treatment of the $z=0$ case), \eqref{A3'} becomes
\begin{multline}
\partial_0 \left( \sqrt{-g} g^{tt} g^{\phi \phi} \partial_0 A_\phi \right)\\ + \partial_r \left(\sqrt{-g} g^{rr} g^{\phi \phi} \partial_r A_{\phi} \right) + z R^{\phi \phi} A_{\phi} = 0.
\end{multline}
Recall from \cite[p.~38]{Catalogue} that the $R_{\phi\phi}$ component of the Ricci tensor is given by
\begin{equation}
R_{\phi \phi} = r^2 \sin^2(\theta) \left(R \ddot{R} + 2 (\dot{R}^2 + k) \right).
\end{equation}
The square root of the metric determinant is given by 
\begin{equation}
\sqrt{-g} = R^3 (t) r^2  \frac{\sin{\theta}}{\sqrt{1-k r^2}}.
\end{equation}
For simplicity, we set $R = 1$ from this point on (for instance, the constancy of $R$ is typically assumed for the current cosmological era). Thereby, raising indices using the metric,
\begin{equation}
R^{\phi \phi} = \frac{R \ddot{R} + 2 (\dot{R}^2 + k)}{R^4 r^2 \sin^{2}(\theta)} =  \frac{2 k}{r^2 \sin^2(\theta)}.
\end{equation}


\noindent
Upon further specifying the ansatz to be
\begin{equation}
A_{\phi} (r, t) = \xi(r) \exp\left(i S(t)  + i S(r) \right)
\end{equation}
with wave vector $K_{\mu}$ such that $K_{t} := \partial_t S(t)$ and $K_{i} := \partial_i S(r)$, one finds that $K_r$, i.e.~the (non-trivial spatial) $r$-component of the wave vector, and $\xi (r)$ are governed by
\begin{widetext}
\begin{multline}\label{vlong}
\partial_t \left(\sqrt{-g} g^{tt} g^{\phi \phi}\right) \left(\partial_t \xi + \xi K_t i\right) + \left(\sqrt{-g} g^{tt} g^{\phi \phi}\right) \left(\partial_t^2 \xi + 2 \left(\partial_t \xi\right) K_t i - \xi K_t^2 + \xi \left(\partial_t K_t\right) i \right) + \\ \partial_r \left(\sqrt{-g} g^{rr} g^{\phi \phi}\right) \left(\partial_r \xi + \xi K_r i\right)+  \left(\sqrt{-g} g^{rr} g^{\phi \phi}\right) \left(\partial_r^2 \xi + 2 \left(\partial_r \xi\right) K_r i  - \xi K_r^2 + \xi \left(\partial_r K_r\right) i\right) + z R^{\phi \phi} \xi = 0.
\end{multline}
\end{widetext}
Exploiting that $\partial_t (\sqrt{-g} g^{tt} g^{\phi \phi}) = 0$ for $R=1$, that $K_t$ is constant, and that $\partial_t \xi(r) = 0$, as well as splitting up the real and imaginary parts of \eqref{vlong}, yields two equations. First, the imaginary part:
\begin{multline}
\partial_r \left(\sqrt{-g} g^{rr} g^{\phi \phi}\right) K_r \xi + \\ \sqrt{-g} g^{rr} g^{\phi \phi} \left( 2 \left(\partial_r \xi\right) K_r + \xi \left(\partial_r K_r\right)\right) = 0
\end{multline}
Upon multiplication by $\xi$, one obtains $\partial_r (\sqrt{-g} g^{rr} K_r \xi^2) = 0$; this is solved by
\begin{equation}
\label{amp}
\xi = \frac{a}{{ \left( \sqrt{-g} g^{rr} g^{\phi \phi} K_r \right)}^{1/2}} ,
\end{equation}
where $a$ is a constant.

For the real part of \eqref{vlong}, we have
\begin{multline}
\sqrt{-g} g^{tt} g^{\phi \phi} \left(\partial_t^2 \xi - \xi K_t^2\right) + \partial_r \left(\sqrt{-g} g^{rr} g^{\phi \phi}\right) \partial_r \xi\\ + \sqrt{-g} g^{rr} g^{\phi \phi} \partial_r^2 \xi - \sqrt{-g} g^{rr} g^{\phi \phi} \xi K_r^2 + z R^{\phi \phi} \xi = 0.
\end{multline}

\noindent
Dividing by $\sqrt{-g} g^{\phi \phi} \xi$, one obtains
\begin{align}
\label{dispersion}
K^{\mu} K_{\mu} &= g^{tt} K_t^2 + g^{rr} K_r^2 \notag \\
&= \frac{z R^{\phi \phi}}{\sqrt{-g} g^{\phi \phi}}  + g^{rr} \frac{\partial_r^2 \xi}{\xi} + \frac{\partial_r (\sqrt{-g} g^{rr} g^{\phi \phi}) } {\sqrt{-g} g^{\phi \phi} } \frac{\partial_r \xi}{\xi},
\end{align}
i.e.~the dispersion relation for the wave. At this stage, note that for the specific constant amplitude ansatz, i.e.~$\xi(r) = \text{const}$, the wave follows null geodesics for $z = 0$ in the sense that the dispersion relation vanishes (in particular, the group velocity becomes unity---this is shown in \cite{AH}), but also that in the case of FLRW spacetimes there is no consistent solution for $K_r$ for $z \neq 0$. (In generic spacetimes, the specific constant amplitude ansatz does not even lead to solutions for $z=0$, as already mentioned above, and as pointed out in \cite{AH2}.)



We will, however, use the solution for $\xi = \xi(r) \neq \text{const}$ obtained in terms of $K_r$ as given by \eqref{amp}; plugging into the dispersion relation \eqref{dispersion} gives
\begin{widetext}
\begin{equation}\label{DispersionFRW}
-K_t^2 + \left(1-k r^2\right) K_r^2=\frac{\left(k r^2-1\right) \partial^2_r K_r}{2 K_r}-\frac{3 \left(k r^2-1\right) (\partial_r K_r)^2}{4 K_r^2} - k \frac{2 + r^2}{4 \left(k r^2-1\right)} + z k \frac{8 k^2 r^4  -16 k r^2  +8 }{4 r^2 \left(1- k r^2\right)^{3/2}},
\end{equation}
\end{widetext}
where $z = 0$ if the additional curvature coupling term is neglected.
Note that even for $z = 0$, the wave exhibits non-trivial dispersion in these coordinates (we discuss further the possibility of coordinate effects below), as long as the amplitude is non-constant. (This is compatible with the finding by \cite{AH2} that in the $z=0$ but \textit{constant} amplitude case the dispersion relation is trivial again.)

Restricting to $S(t) = \omega t$ where $\omega$ is a constant, we now solve numerically (using \textsc{Mathematica}) \eqref{DispersionFRW} in the domain $r \in \left[ 0.1, 0.9\right]$ (in order to avoid issues of coordinate singularities) with $k=1$, subject to the initial conditions $K_r\left(0.1\right)=\omega$ and $K^\prime_r\left(0.1\right)=1$. One could equally well choose $K^\prime_r\left(0.1\right) = 0$, even though this seems to suggest that the wave-vector is position-independent and that a trivial dispersion-relation imposed at one point (as secured via $K_r(0.1) = \omega$) will thereby remain trivial everywhere. This impression, however, is misleading insofar as we are still considering at this point the wave in $\left(t,r\right)$ coordinates, rather than in the $\left( \tau, \rho\right)$ coordinates to be discussed below. (Only setting the analogous initial condition in terms of $\rho$ to $0$, i.e.~requiring $\partial K_{\rho} / \partial \rho =0$ would impose position-independence of the dispersion relation in a physical sense.)
Then, inputting a fixed $r$ in the above domain, one can extract functional relationships between $K_r$ and $\omega$ at different values of $z$. Choosing, for instance, $r=0.2$, one finds the dependencies presented in figure 1.
\begin{figure}
    \centering
    \includegraphics[scale=0.55]{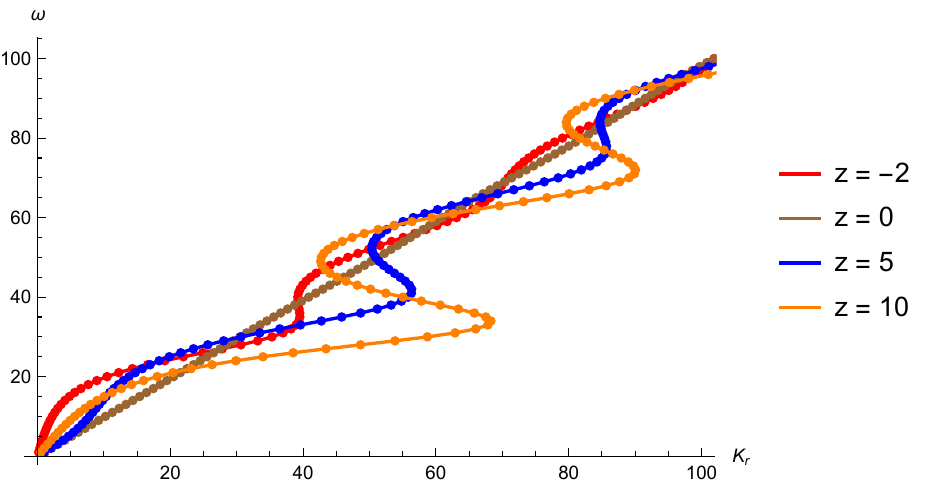}
    \label{fig1}
    \emph{\caption{$\omega = \frac{\partial S(t)}{\partial t} = K_t$ as a function of $K_r$ for a transverse electromagnetic wave in FLRW spacetime ($k = 1$). ($K_r$ is obtained numerically from \eqref{DispersionFRW} with $K_r(0.1)=\omega$, $K_r'(0.1) = 1$, giving rise to the functional relationship between $K_r$ and $\omega$.) Different values of $z$ correspond to the strength of the additional curvature coupling.}}
\end{figure}
From these results, one finds that although even for $z=0$ it is not the case that the group velocity $\partial\omega/\partial K_r$ is exactly unity (where we normalise $c=1$), dispersion effects increase with $z$. Choosing greater $r$ exacerbates the dispersion effects, as one would expect, given the greater radial distance from the point at which the above initial conditions are imposed. From the above results, one also sees clearly that the phase velocity $\omega/K_r$ of the wave is not constant. Finally: if one repeats the calculations of this sections in other spacetimes, such as G\"{o}del, one finds analogous results.

Now, strictly speaking, in order to avoid being misled by coordinate effects when making these judgements on the dispersion relation, one should switch to coordinates that bring the wave equation into a form locally adapted to Minkowski spacetime. Generally, for a spherically symmetric metric given as \cite[p.~4]{AH}
\begin{align}
g_{tt} &= f(t) q(r), \\
g_{rr} &= h(t) b(r), \\ 
g_{\theta \theta} &= h(t) r^2, \\
g_{\phi \phi} &= h(t) r^2 \sin^2(\theta),
\end{align}
the time coordinate $t$ and spatial coordinate $r$ need to be changed, respectively, to \cite[p.~5]{AH}
\begin{align}
\tau &:= \int dt \sqrt{-\frac{f}{h}} ,\\
\rho &:= \int dr \sqrt{\frac{b}{q}} .
\end{align}
Thus, for FLRW spacetime with $R(t)=1$ and $k > 0$,
\begin{align}
\tau &= t , \\
\rho &= \int dr \sqrt{\frac{1}{1-k r^2}} = \frac{1}{\sqrt{k} \sin{(\sqrt{k} r)}} , \label{eq31}
\end{align}
setting all integration constants to zero. Inverting \eqref{eq31}, one obtains
\begin{equation}\label{rrho}
r = \frac{1}{\sqrt{k}} \arcsin{\left(\frac{1}{\sqrt{k} \rho}\right)}.
\end{equation}

This being said, in the above we work in $\left(t,r\right)$ coordinates, for two reasons. First, the form of the coordinate transformation between $r$ and $\rho$, i.e.~\eqref{rrho}, leads to oscillatory functions in the transformed version of \eqref{DispersionFRW}; and unlike the $\left(t,r\right)$ coordinates, the coordinates locally adapted to Minkowski spacetime can only be used to cover FLRW spacetime for relatively small regions as the transformation rule from $r$ to $\rho$ is periodic and thus non-monotonous. Second, given the particular form of \eqref{rrho}, one can be assured that tuning $z$ to different values will lead to different physical effects even in the $\left( \tau, \rho\right)$ coordinates, as the $z$-dependent term of \eqref{DispersionFRW} is non-constant in $\rho$ and not merely in $r$ (this follows from plugging \eqref{rrho} into the $z$-dependent term). Note finally that transforming to $\left( \tau, \rho\right)$ coordinates implies that propagating electromagnetic waves can have dispersion even in familiar cases of e.g.~Schwarzschild spacetimes (and even having set $z=0$; see \cite[\S2]{AH}).

Thus, the above results corroborate the claims made in \cite{AH, AH2} that, outside of the geometrical optics regime, light does not invariably propagate on null geodesics in general relativistic universes; they also make clear the tangible physical effects of including extra curvature couplings in Maxwell's equations that potentially arise due to the ambiguities of the minimal coupling scheme (and thus prior to any semi-classical corrections from QED, as for instance accounted for by the Drummond-Hathrell action, or other sorts of matter couplings \cite{Hertzberg}). There are, however, certain other claims made in \cite{AH} which are questionable: these regard the consistency of the geometrical optics limit. It is to these issues that we now turn; ultimately, we will reject these claims---this constitutes the negative part of our discussion of \cite{AH, AH2}.

\section{Spacetimes with inconsistent geometrical optics limits?}\label{s4}

In \cite{AH}, the authors suggest that certain wave solutions in Gödel spacetime do not have a well-defined geometrical optics limit; that is, no exact wave solutions are consistent with $K^{\mu} K_{\mu} = 0$.
They claim that there is therefore no sense in which these electromagnetic waves can be said to move on null geodesics at high frequencies.



These claims deserve greater scrutiny. To this end, recall the reasoning of \cite{AH} in more detail. For G\"{o}del spacetime, the dispersion relation following from \eqref{A2'} is
\begin{equation}\label{DispersionGoedel}
-\omega^2 + K_x^2 = -\frac{\Omega^2}{2} + \frac{3 K^{\prime 2}_x}{4 K_x^2} - \frac{K^{\prime \prime}_x}{2 K_x},\end{equation}
where $' \equiv \partial_x$. In \cite{AH}, the geometrical optics limit is referred to as the `high-frequency limit', which one can take to be the limit $\omega \rightarrow \infty$; it is then claimed that the limit is successfully taken only if $K^\mu K_\mu \rightarrow 0$.
Note, though, that the R.H.S.~of \eqref{DispersionGoedel} is unequal to zero under $\omega \rightarrow \infty$ and is therefore inconsistent with this requirement. The reason for this is not only due to the presence of the $\Omega \neq 0$ term---as indicated in \cite{AH}---but also due to the presence of the terms in $K'_x$ and $K''_x$, neither of which need vanish when $\omega \rightarrow \infty$. To be fair to \cite{AH}, these latter two terms do vanish when amplitude variations are taken to be negligible (a feature of the geometrical optics limit as discussed at \cite[\S22.5]{MTW} and cited in this context by \cite{AH}), but it is worth recognising that this does not follow from the condition $\omega \rightarrow \infty$ alone. Regardless of whether one takes this into account, however, the point made in \cite{AH}---that terms on the R.H.S.~of \eqref{DispersionGoedel} spoil the consistency of this limit---still stands.
This, indeed, is the origin of the claim in \cite{AH} that the geometrical optics limit is inconsistent in such spacetimes.

Is it, however, correct to understand the geometrical optics regime in terms of a high frequency (and potentially also constant amplitude) limit? Recall from \cite[p.~571]{MTW} that in the geometrical optics limit (traditionally construed) one \textit{also} requires that wavelength be much less than the characteristic scale of curvature variations (in addition to taking amplitude variations to be negligible---an assumption which we will continue to make in this paragraph). Now, in G\"{o}del spacetime, one has that $G_{\mu\nu}u^\mu u^\nu = \Omega^2$, where $G_{\mu\nu}$ is the Einstein tensor and $u^\mu$ is some normalised four-velocity vector \cite[p.~168]{HE}; thus, we see that $\Omega$ is directly related to curvature effects, and so, in the geometrical optics limit, it makes sense on physical grounds to impose two conditions: $\omega \rightarrow \infty$ and $\Omega \rightarrow 0$; alternatively---but not equivalently---one could take the limit $\Omega / \omega \rightarrow 0$. (Whether these two approaches to the geometrical optics limit are indeed equivalent depends upon the convergent series which one selects in the latter: our thanks to Sam Fletcher for pointing this out to us.) In the remainder of this section, we will for concreteness focus on the limit $\Omega / \omega \rightarrow 0$. Taking this limit, one finds from \eqref{DispersionGoedel} that the geometrical optics limit can be taken consistently, and that light does propagate on null geodesics in this limit. (An aside: taking a limit of this kind, involving a comparison of scales, seems to us a promising way of understanding the content of the equivalence principle as it is used in physical practice---see e.g.~\cite{Will}.)

One might take the general point here to be this: a limiting procedure on a physical quantity can be regarded as being physical only if formulated in relation to some other physical quantity. (Compare, for example, the by-now suspect $c \rightarrow \infty$ Newtonian limit in special relativity with the better-accepted $v/c \rightarrow 0$ limit in this theory.) On these grounds, one might claim that the $\omega \rightarrow \infty$ limit deployed in \cite{AH} is not physically justified. While roughly correct, this way of putting things remains a little too quick, for one can now ask: if, when considering some particular limit, there is a sequence of models shown to converge in some well-defined sense, and all of the models are interpreted (i.e., represent some physical possibility), what more could be needed in declaring the limit to be `physical'? Such certainly seems to be the case for the $\omega \rightarrow \infty$ limit taken in \cite{AH}, in which case it is not clear that `physicality' is the most appropriate way in which to couch the issue. Indeed, we do not deny that studying the $\omega \rightarrow\infty$ limit can advance our understanding of the behaviour of light in curved spacetimes: there is nothing \emph{per se} wrong with the study of high frequencies in such spacetimes. Rather, the issue with the limit taken in \cite{AH}, as we see it, is that it does not represent what the authors purport it to represent: in particular, whether `physical' or not, and whether physically illuminating in certain respects or not, it nevertheless does not represent the geometrical optics limit as traditionally understood. (Again, our thanks to Sam Fletcher for discussions on these matters: see \cite{Fletcher2} for related discussions of limiting relations in spacetime theories.)

For maximal clarity, one might introduce a parameter tracking the frequency orders, as done in \cite[\S22.5]{MTW}. It will then be clear that only terms in the equations of motion involving at least two derivatives in the vector potential term $A^\mu$ can, upon a wave ansatz, contribute at the relevant order to the dispersion relation; at the frequency order two, we have $K^{\mu} K_{\mu} = 0$.



The following observation is also worthy of mention. Consider \eqref{DispersionFRW} in FLRW spacetimes.
As long as $k \neq 0$ (as discussed before, non-constant amplitude waves are solutions in flat FLRW---they do not arise only in the geometrical-optical limit approximation), the R.H.S.~is unequal to zero under $\omega \rightarrow \infty$ independently of the chosen curvature coupling (i.e.,~value of $z$) for Maxwell's equations. Therefore, non-flat FLRW spacetime---which, of course, builds the modelling foundation for the standard model of cosmology---would represent another physical example of there being exact wave solutions in spacetimes which (i) do not propagate on null geodesics, and which (ii) do not have a well-defined geometrical optics limit. But again, our same reservations as above apply: if one takes what is arguably the correct geometrical optics limit in this case, which here would effectively involve taking $k \rightarrow 0$ (or $k/\omega\rightarrow 0$), one recovers both the consistency of the limit and the null propagation of light.

In \cite{AH}, approximations are invoked in order to derive from \eqref{DispersionGoedel} the following results for the phase and group velocities of light in G\"{o}del spacetime in the case $\Omega x \ll 1$:
\begin{align}
    v_{\text{ph}} &= \frac{\omega}{K_x} \approx 1 + \frac{1}{2}\Omega^2 x^2 , \label{vph} \\
    v_{\text{gp}} &= \frac{\partial \omega}{\partial K_x} \approx 1 + \frac{1}{2}\Omega^2 x^2. \label{vgp}
\end{align}
This seems to corroborate the claim that, in the $\omega\rightarrow \infty$ limit, light still does not propagate on null geodesics. However, there are two points to note here. First: if one also takes $\Omega\rightarrow 0$, then null propagation is retained (the authors of \cite{AH} note this in the context of flat spacetimes, but not in the context of the understanding of the geometrical optics limit discussed above). Second: one might wonder whether \eqref{vph} and \eqref{vgp} are artefacts of truncating a series too early.
Continuing the expansion in \eqref{vph} to higher order, one finds that
\begin{widetext}
\begin{multline}\label{Kexpand}
K_x = \omega \left(1 -\frac{x^2 \Omega ^2}{2} +\frac{x^4 \Omega ^4}{6} -\frac{17 x^6 \Omega ^6}{360} +  \cdots \right) +\omega^3 \left(\frac{x^4 \Omega ^2}{6} -\frac{13 x^6 \Omega ^4}{90} + \frac{8 x^8 \Omega^6}{105} - \frac{899 x^{10} \Omega^{8}}{28350} + \cdots \right) \\ + \omega^5 \left(-\frac{x^6 \Omega ^2}{45} + \frac{x^8 \Omega ^4}{21} - \frac{1699 x^{10} \Omega ^6}{37800} + \frac{108593 x^{12} \Omega ^8}{3742200} - \cdots \right) + \cdots
\end{multline}
\end{widetext}
Noting the contribution of the additional terms in \eqref{Kexpand}, it is not at all clear that \eqref{vph} and \eqref{vgp} are the final story on the phase and group velocities of light in G\"{o}del spacetime in the case $\Omega x \ll 1$.


To summarise: in \S\ref{s3} we fortified the claims made in \cite{AH, AH2} that light, outside of the geometrical optics limit, need not propagate on null geodesics. However, in this section we have called into question the claims made in those articles that this limit is inconsistent in spacetimes such as G\"{o}del.

\section{Propagation speed in curved spacetime}\label{s5}

If one accepts that light need not propagate along null geodesics when properly modelled by solving Maxwell's equations in curved spacetimes, then one decouples light propagation from null cone structure. In \cite{Hojman}, Hojman accordingly takes the above-discussed results regarding the varying group velocity of light in curved spacetimes as motivation for redefining the proper time interval: the proper time interval is to reflect the actual interaction of light with the gravitational field (as represented by the background metric) instead of just the idealised interaction of light with the gravitational field in the geometrical optics limit (in which light indeed moves on null geodesics). In response to this, however, one might argue that it is not clear that null cone structure is most sensibly tied to light structure to begin with: rather, the proper time interval might better be understood through the major conceptual roles of encoding causal structure, and quantifying distance between path-connectible points alone.

More significantly: the causal significance of the proper time interval is not necessarily endangered by the mere finding that there are superluminal group velocities---as for instance observed from the dispersion relations above for Gödel and FLRW spacetimes---since superluminal propagation is not linked straightforwardly to superluminal \emph{signalling}. After all, signal propagation speed (the propagation relevant for causal considerations) is bounded by the front velocity which, for a general set of PDEs \cite[p.~15]{Shore}, is given by \citep{LIM, Shore}
\begin{equation}
    v_{\text{wf}} = \lim_{\omega\rightarrow\infty}\frac{\omega}{|\vec{k}|} = \lim_{\omega\rightarrow\infty}v_{\text{ph}}\left(\omega\right) .
\end{equation}

\noindent
In fact, the front velocity can be shown to be unity in curved spacetime for a wide class of second-order tensorial PDEs, including \eqref{A2} and \eqref{A3} \cite[\S3.2]{Friedlander}, in which case the upper bound to signalling speed indeed seems to be unity, as argued in the classic work of \cite{Brillouin} (see \cite{Milonni} for further discussion). That, of course, does not mean that this bound is always saturated. In fact, it is generally accepted that in normal dispersive media, such as water, causal influences propagate at velocities less than the front velocity (cf.~\cite{CausalLoop}); furthermore, it has been shown in the seminal work \cite{Nature} that in anomalous dispersive media causal influences indeed propagate at speeds slower than unity (despite a faster-than-unity group velocity). 


Similarly, then, the instantaneous signal speed of an electromagnetic wave in curved spacetime---the curvature of spacetime mimics across different frequencies and curvature scales the effects of both normal and anomalous dispersive media---might be seen generally to deviate downwards from unity. (Such an effect can also be argued for on the basis of the tail problem, as discussed above.) The precise amount of downwards deviation will be a function of the curvature length scale relative to frequency at which the wave is considered. 

Now, upon taking the idealising geometrical optics limit, the causal structure will still be given by the light cone structure as long as at most corrections to the usual Maxwell equations \eqref{A2} (or, for what it is worth, \eqref{A3}) are added that do \textit{not} involve terms of equal to or higher order in $(\partial A)^2$ or $\partial \partial A$ \cite[\S3.1]{Hertzberg}. When terms of this form are added to the Maxwellian equations of motion (as for instance the case when taking into account semi-classical corrections from QED), the geometrical optical limit generally results in a non-zero dispersion relation (in the case of the QED corrections this is the already-discussed Dummond-Hathrell effect). The impact of the aforementioned terms on the dispersion relation in the geometrical optics limit is easy to understand from the fact that any derivative on $A^\mu$ amounts, upon plugging in the wave ansatz, to pulling down a frequency factor $\omega$. Thus, terms of the described form are of order $\omega^2$ and cannot be neglected upon the geometrical-optical limit: see \cite{Hertzberg} for a discussion.

This being said, even in the more general case in which these terms are included, Hertzberg and Sandora \cite[\S4.2]{Hertzberg} have identified a back-reaction mechanism such that in the context of general relativity (albeit not for curved spacetimes in general) superluminal signalling can occur only at scales smaller than a photon's wavelength but is otherwise averaged out; given that superluminal signalling events for a photon only endure at lengths scale smaller than its own wavelength, such signalling might well be regarded as being operationally meaningless.

Outside of the regime of the geometrical optics limit, a light clock realised through periodic bouncing of a light signal (so-called `Langevin clocks') will in practice not read out reliably the proper time along its worldline interval. But, to repeat: on taking the idealising geometrical optics limit and setting aside the above-discussed additional coupling terms, contra \cite{clocks}, an idealised light clock construction modelled by Maxwellian theory in curved spacetime will still read out the proper time interval along its worldline (for, upon idealisation, the signal speed is unity, and the assumptions of \cite{Fletcher} obtain). Moreover, even when corrections from QED are taken into account, the aforementioned mechanism \emph{à la} Hertzberg and Sandora guarantees that deviations from the light cone structure are minuscule, or even operationally meaningless, in the context of general relativity. In the context of \cite{AH, AH2}, our point in this section is, ultimately, a simple one: even granting the claims made in those papers, it is not at all clear that superluminal \emph{signalling} is implicated.

\section{Discussion}

The finding of \cite{AH}---that there are electromagnetic wave solutions in Gödel spacetimes not amenable to the geometrical optics limit---seems to stand in tension with other major results in the literature on electromagnetic wave propagation in curved spacetimes, as were mentioned in the introduction: (i) Friedlander shows that the front velocity of waves in curved spacetime is always 1 (see \cite[Thm.~3.2.1]{Friedlander}); (ii) in \cite{GW}, the authors demonstrate the small bodies constructed from Maxwell fields in curved spacetimes `track' (in a technical sense) null geodesics, and claim that this is a special case of the geometrical optics limit. The present article should ameliorate any such tension, for we have demonstrated that, given a physical understanding of the geometrical optics limit, this limit can invariably be taken consistently, that light invariably propagates along null geodesics in this limit, that in particular the front velocity is unity, and that, nevertheless, the signal velocity may be less than unity outside of the geometrical optics limit.


\section*{Acknowledgements}

We are grateful to Jacob Barandes, Juliusz Doboszewski, John Dougherty, Sean Gryb, Dennis Lehmkuhl, Tushar Menon, Erhard Scholz, Nic Teh, and Jim Weatherall for valuable discussions, and to Sam Fletcher, Chris Smeenk, and the three anonymous referees for very helpful feedback on previous drafts. N.L. acknowledges financial support from the John Templeton Foundation (grant 61048).

\bibliographystyle{plain}

\end{document}